\documentclass[conference]{IEEEtran}

\IEEEoverridecommandlockouts
\usepackage{cite}
\usepackage{amsmath,amssymb,amsfonts}
\usepackage{algorithmic}
\usepackage{graphicx}
\usepackage{textcomp}
\usepackage{xcolor}
\def\BibTeX{{\rm B\kern-.05em{\sc i\kern-.025em b}\kern-.08em
    T\kern-.1667em\lower.7ex\hbox{E}\kern-.125emX}}

\usepackage{fancyhdr}
\begin{document}

\title{SAPIEN: Affective Virtual Agents Powered by Large Language Models*\\
\thanks{NSF and NSF REU IIS-1750380, Seedling from Goergen Institute for Data Science (GIDS), and  Gordon and Moore Foundation.}
}

\author{\IEEEauthorblockN{Masum Hasan\IEEEauthorrefmark{1},
Cengiz Ozel\IEEEauthorrefmark{2}, Sammy Potter\IEEEauthorrefmark{3} and
Ehsan Hoque\IEEEauthorrefmark{4}}
\IEEEauthorblockA{Department of Computer Science,
University of Rochester\\
Rochester, NY, United States\\
Email: \{\IEEEauthorrefmark{1}m.hasan@,
\IEEEauthorrefmark{2}cozel@cs.,
\IEEEauthorrefmark{3}spotter14@u.,
\IEEEauthorrefmark{4}mehoque@cs.\} rochester.edu
}}

\maketitle
\thispagestyle{fancy}

\begin{abstract}
In this demo paper, we introduce SAPIEN, a platform for high-fidelity virtual agents driven by large language models that can hold open domain conversations with users in 13 different languages, and display emotions through facial expressions and voice. The platform allows users to customize their virtual agent's personality, background, and conversation premise, thus providing a rich, immersive interaction experience. Furthermore, after the virtual meeting, the user can choose to get the conversation analyzed and receive actionable feedback on their communication skills. This paper illustrates an overview of the platform and discusses the various application domains of this technology, ranging from entertainment to mental health, communication training, language learning, education, healthcare, and beyond. Additionally, we consider the ethical implications of such realistic virtual agent representations and the potential challenges in ensuring responsible use.


\end{abstract}

\begin{IEEEkeywords}
Virtual Avatars, Virtual Agents, Affective AI, Large Language Models
\end{IEEEkeywords}

\section{Introduction}


Allowing a user to define the traits and characteristics of a virtual agent, carrying a dynamic conversation, and receiving automated feedback has been an open-ended research problem for many years \cite{Hoque-sensing}. The rapid advancement of Large Language Models (LLMs) in recent years has enabled  possibilities in designing user experiences that didn't exist before \cite{chatgpt, cluaude, bard}. In this demo, we present Synthetic Anthropomorphic Personal Interaction ENgine (SAPIEN), a platform for LLM-powered high-fidelity virtual agents that can engage in real-time open-domain conversations, while also expressing emotions through voice and facial expressions.

One of the notable features of SAPIEN is its extensive range of customization options, allowing users to engage in immersive and meaningful interactions. Users can choose from a wide range of virtual agent avatars that reflect a diverse array of ages, gender, and ethnicities. Going further, users can select the desired personality, background, and conversational context of a virtual agent, creating an experience tailored to their specific needs or preferences.

Once a virtual agent is selected and its traits are defined, users can begin a real-time video call interaction with it. With the help of the large language model, the virtual agents dynamically adjust their emotional state, vocal, and facial expressions, showcasing a spectrum of seven basic emotions.

SAPIEN leverages state-of-the-art models in Speech-to-Text \cite{speech2text, speech2text2}, Text-to-Speech \cite{text2speech, text2speech2, text2speech3}, and large language modeling \cite{bard, chatgpt, alpaca, gpt3, gpt4, instruct-gpt, openassistant}. The virtual agents fluently speak thirteen different languages and counting, making it accessible across a global user base.

\begin{figure}
    \centering
    \includegraphics[width=\linewidth]{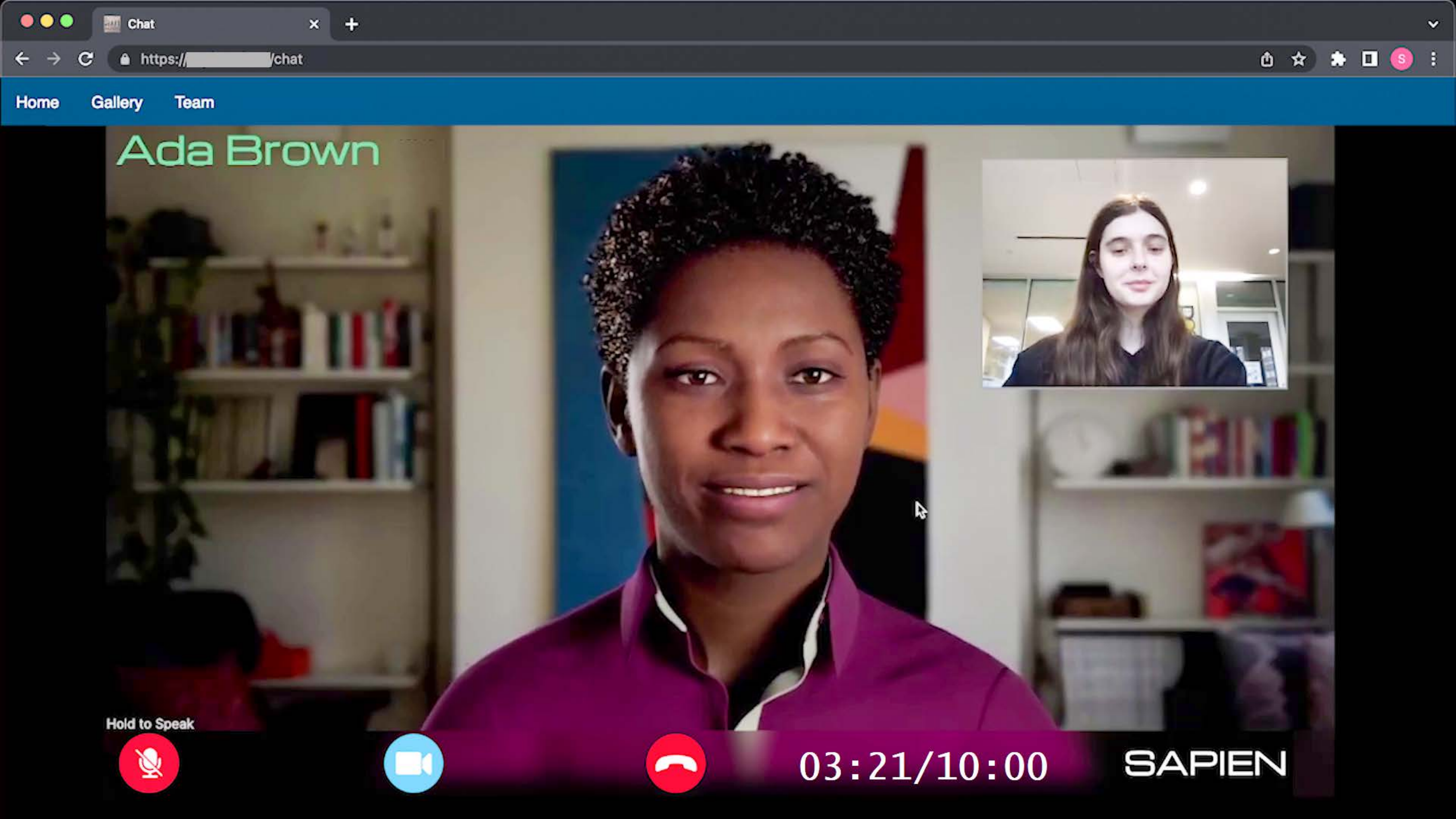}
    \caption{Face-to-face video call interaction with SAPIEN\textsuperscript{TM} Virtual Agent}
    \label{fig:enter-label}
\end{figure} 

\begin{figure*}
    \centering
    \includegraphics[width=0.9\linewidth]{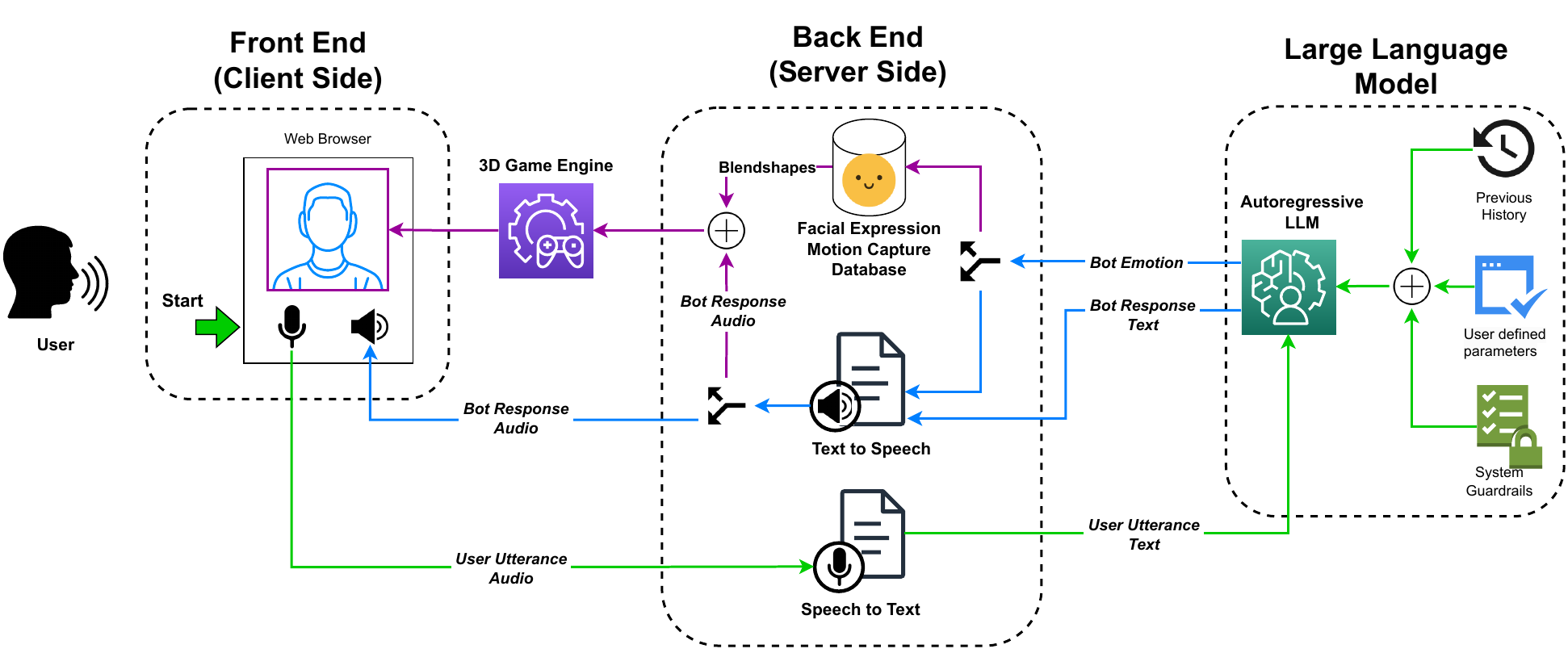}
    \caption{A single turn conversation flow in SAPIEN. User utterance is transcribed and sent to LLM. The LLM response is spoken out by the virtual agent.}
    \label{fig:flow}
\end{figure*} 

Upon finishing a video call with the virtual agents, a user can choose to get their conversation analyzed for personalized feedback. The system provides AI-generated feedback to the user based on the user's goal. The user can decide the topic of the feedback to suit their learning goal and repeat the conversation until the learning goal is met. The inherent flexibility of the virtual agent persona and the feedback could make it potentially applicable to a myriad of applications, including communication training, language learning, and professional applications like healthcare, sales, and leadership training. 

With the rising technical capabilities of LLMs, there is expected to be a drastic shift in the labor market in the coming years \cite{gpt-jobs}. According to recent studies \cite{gpt-jobs}, the importance of the job market is going to shift from hard technical skills to soft ``human'' skills. In this changing landscape, SAPIEN can help people adapt and cope, by helping them cultivate human skills with the help of AI.

\section{System Description}


The overall working of SAPIEN Virtual Agents, referred to as `Bot' for simplicity, is represented in Figure \ref{fig:flow}. The SAPIEN system is initialized when a user's speech utterance is captured and transmitted to our back-end server for processing. This utterance is transcribed into text by a high-precision Speech to Text (STT) model \cite{speech2text, speech2text2, speech2text3, speech2text4} and subsequently processed by an autoregressive Large Language Model (LLM) fine-tuned for instruction following \cite{gpt3, instruct-gpt, gpt4, bard, cluaude, vicuna, openassistant, alpaca}.

The LLM is conditioned on user-defined parameters like personality traits, conversation premise, user information, and previous conversation history. To prevent inappropriate or offensive behavior, the LLM also adheres to system guardrails. A notable aspect of the LLM is also predicting the virtual agent's emotional state. Conditioning on the user-defined parameters, system guardrails, and previous conversation history, the LLM is instructed to generate the bot's response, alongside the appropriate emotional state of the bot from the following list: Neutral, Happy, Sad, Angry, Surprised, Afraid, and Disgusted.

This emotional state, along with the text response, is used to generate an audio file of the bot's response using a Text to Speech (TTS) model. Concurrently, the emotional state triggers the selection of a corresponding facial expression from our pre-recorded motion capture database. 
This facial expression data, in the form of blendshapes, is passed to a 3D game engine to animate the virtual agent.

The resultant animation and generated audio are synchronized, forming a coherent, visually expressive response from the virtual agent. This combined output is streamed to the user's web browser in near real-time, allowing for an immersive experience close to an actual video call.

Once the conversation is over, the user can opt-in to receive feedback on their conversation. An LLM is instructed to analyze the conversation transcript based on the user's goal, identify strengths and weaknesses on the user's communication skill, and generate actionable feedback for the user.
\section{Applications}



The customizability of the conversation scenario, dynamic dialogues, and the feedback system combined make SAPIEN uniquely suitable for a variety of communication training purposes. For example, the system can be used as a communication practice tool for people with social anxiety or neurodiversity \cite{roc-hci-autism, lissa}, public speaking \cite{roc-speak}, job interviews \cite{Hoque-MACH}, helping elderly with social skills \cite{Razavi-TIIS-Autism}, and even speed dating \cite{Ali-ACII-LISSA}. It also has an excellent potential for professional applications. Such as training doctors in bedside manners or delivering difficult news to their patients \cite{sophie}, personalized training for leadership, business negotiation, sales, marketing, etc. The multilingual ability makes the platform a powerful tool for language learners. Furthermore, the non-judgemental, low stake, repeatable conversations with virtual agents make the platform a helpful tool for anyone to roleplay any difficult interpersonal scenario in a personal or professional setup.
\section{The Demo}

Our platform is hosted in the cloud and accessible from any part of the world. During the conference demo, we wish to have the visitors live interact with SAPIEN virtual agents in a variety of interesting scenarios and receive immediate feedback on their communication skills. We will also prepare some pre-recorded user interaction videos to demonstrate any rare or difficult cases or as a backup for technical failures.
\section*{Ethical Impact Statement}


SAPIEN is designed to augment and enrich our capacity for communication, empathy, and understanding, but not substitute human connections. To safeguard against potential emotional dependencies on the system, SAPIEN does not retain the memory of previous interactions, and the conversations are limited to a 10 minutes window with a warning at the 8-minute mark. To prevent the practice of bullying or abusive behaviors using our system, we enabled our virtual agents to end the video call if the user repeatedly displays aggressive or offensive behavior. We are continuously investigating more safety and ethical issues regarding the use of the system.

\bibliographystyle{IEEEtran}
\bibliography{ref}

\end{document}